\documentclass[prl,noshowpacs,english,superscriptaddress,twocolumn]{revtex4}
\usepackage{amsmath,amssymb,amsfonts,mathrsfs}
\usepackage{amsthm}
\usepackage{CJK}
\usepackage{bm}
\usepackage{babel}
\usepackage{graphicx}
\allowdisplaybreaks
\usepackage{braket}
\usepackage[breaklinks]{hyperref}
\usepackage{color}
\hypersetup{colorlinks=true, linkcolor=blue, citecolor=blue, filecolor=blue, urlcolor=blue}

\begin{document}

\title{Intrinsic quantum anomalous Hall phase induced by proximity in germanene/Cr$_2$Ge$_2$Te$_6$ van der Waals heterostructure}

\author{Runling Zou}
\affiliation{Institute for Structure and Function $\&$ Department of Physics, Chongqing University, Chongqing 400044, P. R. China}

\author{Fangyang Zhan}
\affiliation{Institute for Structure and Function $\&$ Department of Physics, Chongqing University, Chongqing 400044, P. R. China}

\author{Baobing Zheng}
\affiliation{College of Physics and Optoelectronic Technology, Nonlinear Research Institute, Baoji University of Arts and Sciences, Baoji 721016, P. R. China}

\author{Xiaozhi Wu}
\affiliation{Institute for Structure and Function $\&$ Department of Physics, Chongqing University, Chongqing 400044, P. R. China}

\author{Jing Fan}
\affiliation{Center for Computational Science and Engineering, Southern University of Science and Technology, Shenzhen 518055, P. R. China}

\author{Rui Wang}
\email{rcwang@cqu.edu.cn}
\affiliation{Institute for Structure and Function $\&$ Department of Physics, Chongqing University, Chongqing 400044, P. R. China}

\begin{abstract}
A van der Waals heterostructure combined with intrinsic magnetism and topological orders have recently paved attractive avenues to realize quantum anomalous Hall effects. In this work, using first-principles calculations and effective model analysis, we propose that the robust quantum anomalous Hall states with sizable band gaps emerge in the van der Waals heterostructure of germanene/Cr$_2$Ge$_2$Te$_6$. This heterostructure possesses high thermodynamic stability, thus facilitating its experimental fabrication. Furthermore, we uncover that the proximity effect enhances the coupling between the germanene and Cr$_2$Ge$_2$Te$_6$ layers, inducing the nontrivial band gaps in a wide range from 29 meV to 72 meV. The chiral edge states inside the band gap, leading to Hall conductance quantized to $-e^2/h$, are clearly visible. This findings provide an ideal candidate to detect the quantum anomalous Hall states and realize further applications to nontrivial quantum transport at a high temperature.
\end{abstract}



\maketitle
Due to the coexistence of electronic band topology and magnetic orders, the design and exploration of exotic topological quantum states in magnetic materials attract intensive interest in condensed matter physics as well as materials science. Among these topological phases of matter, the quantum anomalous Hall (QAH) effect is one of the most fascinating topics. The QAH insulators (i.e., Chern insulators) possess dissipationless chiral edges states characterized by integer Chern numbers $\mathcal{C}$, resulting in quantized Hall conductance in units of $e^2/h$ \cite{PhysRevLett.49.405, Haldane1988}. Topological properties in QAH insulators, induced by spontaneous magnetization without external magnetic fields, provide a reliable avenue to realize next-generation electronic devices with ultralow power dissipation. Therefore, extensive studies, including theoretical and experimental proposals, have always been carried out to explore candidates with QAH effects \cite{Yu2010,PhysRevB.85.045445,Fang2014,PhysRevLett.115.186802,PhysRevLett.119.026402,PhysRevB.92.165418,PhysRevB.97.085401,PhysRevLett.110.116802, Chang2013,PhysRevLett.114.187201, PhysRevLett.113.137201,Natphys2014}. The QAH effect was firstly realized in  magnetically doped topological insulators (TIs), i.e., Cr-doped (Bi,Sb)$_2$Te$_3$ thin films \cite{Chang2013,PhysRevLett.114.187201, PhysRevLett.113.137201,Natphys2014}. But, unfortunately, disorders of random magnetic dopants in doped magnetic systems do not prefer to form ordered magnetic structures. Therefore, the laboratory synthesis of these doped magnetic TIs is quite challenging, leading to important quantum phenomena only observed at ultralow temperatures. In comparison with the doped magnetic systems, a pristine crystal is in favor of forming long-rang magnetic orders. Thus, the realization of intrinsic QAH effects facilitates their realistic applications at high temperatures. However, due to the lack of suitable candidates, the experimental observation of QAH effects in intrinsic magnetic materials is relatively slow even though there are lots of previously predicted QAH systems \cite{PhysRevB.85.045445,PhysRevLett.115.186802,PhysRevLett.119.026402,PhysRevB.92.165418,PhysRevB.97.085401,PhysRevLett.110.196801,PhysRevLett.116.096601,PhysRevB.98.245127}.

Very recently, breakthroughs of intrinsic QAH effects have been achieved in a van der Waals (vdW) layered magnetic TI MnBi$_2$Te$_4$ \cite{PhysRevLett.123.096401,PhysRevX.9.041038,Lieaaw5685,NC2020,Natmater2020,Dengeaax8156}. The compound MnBi$_2$Te$_4$ is stacked in a configuration of Te-Bi-Te-Mn-Te-Bi-Te septuple layers. Significantly, the evidence of zero-field QAH effects has been observed in a five-septuple-layer of MnBi$_2$Te$_4$ up to 1.4 K by Deng \textit{et.al} \cite{Dengeaax8156}. Beyond MnBi$_2$Te$_4$, other vdW superlattice-like Mn-Bi-Te structures with tunable chemical formulas as (MnBi$_2$Te$_4$)$_m$(Bi$_2$Te$_3$)$_n$ (where $m$ and $n$ are integer) have also been proposed \cite{PhysRevLett.123.096401,Wueaax9989,PhysRevX.9.041039,PhysRevB.100.155144}. Through tuning numbers of layers, stacking configurations, or compositions, these vdW heterostructures have been further proposed to investigate rich topological quantum phenomena between magnetism and topology, such as QAH phases \cite{PhysRevLett.123.096401,Lieaaw5685,Dengeaax8156}, axion insulator states \cite{Natmater2020}, antiferromagnetic TIs \cite{NC2020,Natmater2020}, and their phase transitions \cite{PhysRevLett.123.096401}.

These advances mentioned above forward a strategy for investigating magnetic topological phases and especially QAH effects in vdW layered materials. Inspired by the significant achievement in vdW Mn-Bi-Te systems, it is desirable to find other two-dimensional (2D) vdW heterostructures with intrinsic magnetism incorporating topological orders. However, as is well known, the long-rang magnetic order in 2D systems is often strongly suppressed by thermal fluctuations \cite{PhysRevLett.17.1133}, and thus 2D magnets are rarely reported in literatures. Recently, the growth technology for 2D intrinsic ferromagnetic (FM) semiconductors has achieved important progress, and atomic layers of CrI$_3$ \cite{Nature2017cri3} and Cr$_2$Ge$_2$Te$_6$ \cite{Naturecrgete} were realized by mechanical exfoliation. These 2D ferromagnets can potentially act as ideal basic units for designing 2D vdW heterostructures with FM orders driven by proximity effects \cite{PhysRevLett.123.016804, Nano4567,Houeaaw1874}. With this guidance, the QAH effects in graphene-based heterostructures were predicted \cite{PhysRevB.92.165418,PhysRevB.97.085401}, but nontrivial band gaps of these heterostructures are very small due to the extremely weak spin-orbital coupling (SOC) of graphene. This seriously hinders to observe QAH effects at high temperatures. Therefore, the design of promising vdW heterostructures with large nontrivial band gaps, which can also easily be fabricated in the laboratory, is highly desired.

In this work, we demonstrate that the vdW heterostructure germanene/Cr$_2$Ge$_2$Te$_6$ can satisfy the above criteria. Germanene is a well-known time-reversal ($\mathcal{T}$)-preserved QSH insulator, which is a monolayer composed of Ge atoms in a 2D honeycomb lattice with a low-buckled geometry \cite{PhysRevLett.102.236804,PhysRevLett.107.076802}. Based on first-principles calculations, we show that this heterostructure with the out-of-plane FM order hosts high thermodynamic stability that benefits its growth in experiments. In the absence of SOC, the effective model and symmetry analysis reveal that germanene/Cr$_2$Ge$_2$Te$_6$ is a spin-polarized semimetal with a quadratic nodal point. When SOC is considered, the quadratic nodal point shrinks to a sizable band gap of 29 meV. Due to weak vdW interactions, the nontrivial band gap can be tunable up to 72 meV by slightly applying an external pressure. Therefore, our work provides a reliable platform to realize intrinsic QAH effects at high temperatures.

\begin{figure}
	\centering
	\includegraphics[scale=0.21]{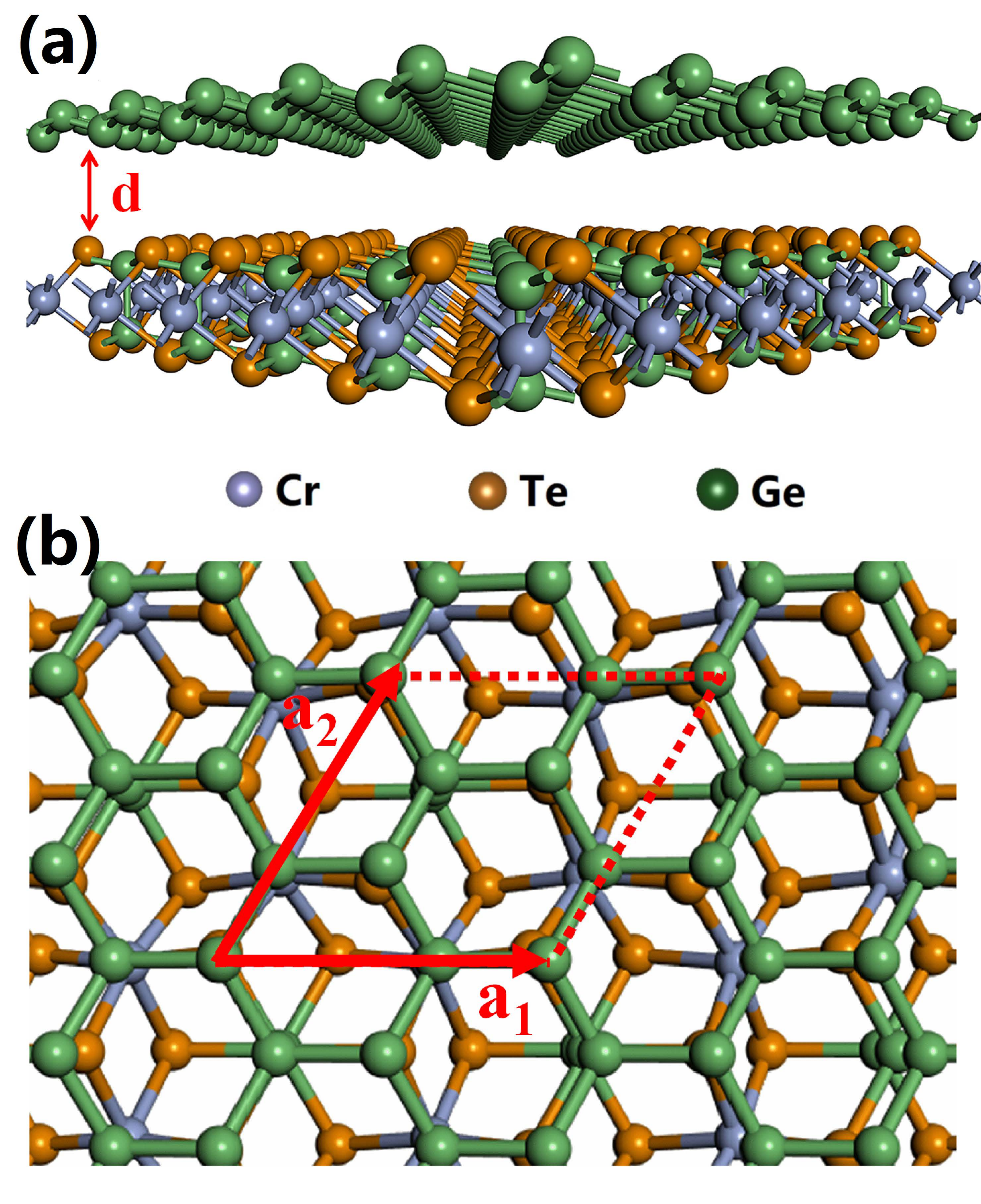}
	\caption{(a) Side and (b) top views of the vdW heterostructure germanene/Cr$_2$Ge$_2$Te$_6$ with the interlayer distance $d$. The hexagonal unit cell is marked by red-lines in (b), which is composed by $p$($\sqrt{3}\times \sqrt{3}$) unit cells of germanene and a $p$($1\times 1$) unit cell of Cr$_2$Ge$_2$Te$_6$.
\label{figure1}}
\end{figure}

To depict electronic and topological properties of the vdW heterostructure germanene/Cr$_2$Ge$_2$Te$_6$, we carried out first-principles calculations based on the density functional theory (DFT) \cite{Hohenberg,Kohn} as implemented in the Vienna Ab initio Simulation Package (VASP) \cite{Kressecom,Kresse2} (see details in the Supplemental Materials (SM) \cite{SM}). As illustrated in Figs. \ref{figure1}(a) and \ref{figure1}(b), we respectively show the side and top views of the germanene/Cr$_2$Ge$_2$Te$_6$ heterostructure, i.e., germanene on top of a single layer of Cr$_2$Ge$_2$Te$_6$. The unit cell of this heterostructure is composed by a $p$($\sqrt{3}\times \sqrt{3}$) unit cells of germanene and a $p$($1\times 1$) unit cell of Cr$_2$Ge$_2$Te$_6$. The lattice mismatch between the germanene and  Cr$_2$Ge$_2$Te$_6$ layers is only 1\%. This low mismatch indicates small strain, which will facilitate the growth in experiments. To obtain the equilibrium geometry of germanene/Cr$_2$Ge$_2$Te$_6$ heterostructure, we optimize the compositive structure starting from several typical stacking configurations. The optimized equilibrium distance is $d_{0}=3.48$ {\AA}, i.e., the interaction between the germanene and Cr$_2$Ge$_2$Te$_6$ layers is a typical vdW bond. From the lowest energy structure that we have obtained, the germanene layer is further moved along the representative $\mathbf{a}_{1}$ and $\mathbf{a}_{1}+\mathbf{a}_{2}$ directions [see Fig. \ref{figure1}(b)] with respect to the Cr$_2$Ge$_2$Te$_6$ layer to make sure that this configuration is really minimum in energy. Moreover, we find that the energy differences among different stacking configurations are quite small; that is, there may be the possibility that the germanene layer is trapped in a local minimum rather than the groundstate while the growth process. Therefore, we calculated electronic properties of several stacking configurations in comparison with that of the equilibrium structure. The results are insensitive to different stacking formalisms, implying that topological features of the germanene/Cr$_2$Ge$_2$Te$_6$ heterostructure are robust [see Fig. S1 in the SM \cite{SM}]. Given the equilibrium structure, the binding energy per Ge atom is calculated by $E_{\mathrm{b}}=(E_{\mathrm{Ge}}+E_{\mathrm{Cr}{_2}\mathrm{Ge}_{2}\mathrm{Te}{_6}}-E_{\mathrm{total}})/N$,
where $E_{\mathrm{Ge}}$, $E_{\mathrm{Cr}{_2}\mathrm{Ge}_{2}\mathrm{Te}{_6}}$, and $E_{\mathrm{total}}$ are the total energies of the isolated germanene layer, isolated Cr$_2$Ge$_2$Te$_6$ layer, and germanene/Cr$_2$Ge$_2$Te$_6$ heterostructure, respectively; and $N=6$ is the number of Ge atoms in one unit cell. We obtain that the binding energy $E_{\mathrm{b}}$ is about 133 meV, indicating that the formation of this heterostructure further improves the thermodynamic stability.

\begin{figure}
	\centering
	\includegraphics[scale=0.40]{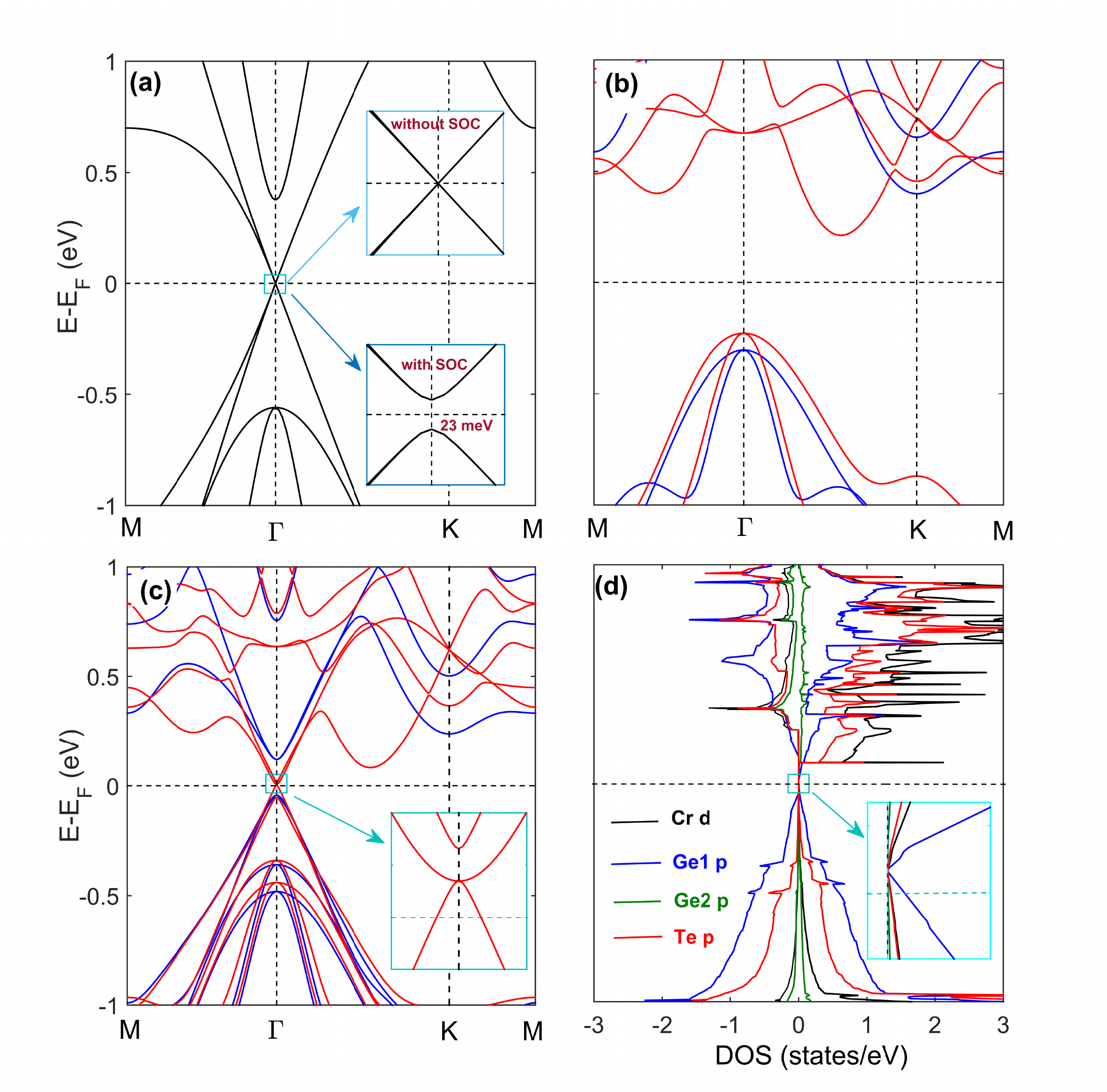}
	\caption{(a) The electronic band structure of $p$($\sqrt{3}\times \sqrt{3}$) germanene. The upper and lower insets are the enlarged views without and with SOC at the $\Gamma$ point, respectively. (b) The spin-polarized electronic band structure of the Cr$_2$Ge$_2$Te$_6$ monolayer in the absence of SOC. (c) The spin-polarized electronic band structure of the vdW germanene/Cr$_2$Ge$_2$Te$_6$ heterostructure in the absence of SOC. The inset exhibits a quadratic nodal point at the $\Gamma$ point. In (b) and (c), the red and blue lines respectively represent the majority and minority spin channels. (d) The spin-resolved partial density of states in the absence of SOC. Ge1 denotes the Ge atoms in the germanene layer, and Ge2 denotes the Ge atoms in the Cr$_2$Ge$_2$Te$_6$ layer.
\label{figure2}}
\end{figure}

The electronic band structure of the pristine $p$($\sqrt{3}\times \sqrt{3}$) germanene is shown in Fig. \ref{figure2}(a). Due to the band folding, the bands at the K point in the case of a germanene unit cell are mapped to the $\Gamma$ point, resulting in a linear Dirac cone at the $\Gamma$ point in the absence of SOC. When the SOC effect is present, germanene is a $\mathbb{Z}_2$ QSH insulator with a nontrivial gap $\sim$23 meV, which agrees well with the previous result \cite{PhysRevLett.107.076802}. The spin-polarized electronic band structure of the Cr$_2$Ge$_2$Te$_6$ monolayer is depicted in Fig. \ref{figure2}(b). We can see that the valence band maximum and conduction band minimum are both contributed by the majority-spin channel, exhibiting a indirect gap. As a result, the Cr$_2$Ge$_2$Te$_6$ monolayer is a FM semiconductor with a saturation magnetic moment of $\sim$ 6 $\mu_{B}$ per unit cell \cite{Naturecrgete}.

Next, we focus on the electronic properties of the vdW germanene/Cr$_2$Ge$_2$Te$_6$ heterostructure. In Fig. \ref{figure2}(c), we show the spin-polarized electronic band structure along high-symmetry directions. The bands in the absence of SOC reveal that the Cr$_2$Ge$_2$Te$_6$ layer completely magnetizes the germanene layer. In this case, this heterostructure is a half-metal; that is, there is a semiconducting feature with a band gap $\sim$163 meV in the minority-spin channel and a semimetallic feature in the majority-spin channel. The one unit cell of the vdW heterostructure has a magnetic moment of $\sim$  6.04 $\mu_{B}$, which is slightly larger than that $\sim$6 $\mu_{B}$ of the isolated Cr$_2$Ge$_2$Te$_6$ layer. In Fig. \ref{figure2}(d), we illustrate the spin-resolved partial density of states in the absence of SOC. It is found that the states around the Fermi level are mainly contributed by the majority-spin channel of Ge1 $p$ orbitals of germanene. Besides, there are also a small amounts of Cr $d$ and Ge2 $p$ orbitals near the Fermi level, implying the presence of hybridization and charge transfer between the germanene and Cr$_2$Ge$_2$Te$_6$ layers. Therefore, the electronic properties of germanene are changed due to the formation of this vdW heterostructure. To determine the behavior of low-energy excitations, we plot the enlarged view of band structures around the $\Gamma$ point [see the inset of Fig. \ref{figure2}(c)]. It is found that there is a nodal point at the $\Gamma$ point. This nodal point is about $\sim$ 7 meV above the Fermi level. That is to say, the proximity makes lightly hole doped in germanene. Furthermore, the dispersion around this nodal point are quadratic, which is different from the linear Dirac Cone of the isolated germanene.

In order to insightfully understand the quadratic dispersions around the $\Gamma$ point, we construct a $2\times 2$ $k\cdot p$ effective Hamiltonian to describe the two crossing bands as
\begin{equation}\label{eqH}
\mathcal{H}(\mathbf{k})=f(\mathbf{k})\sigma_{+}+f(\mathbf{k})^{*}\sigma_{-},
\end{equation}
where $\mathbf{k}$ is the wave vector referenced to the $\Gamma$ point, $f(\mathbf{q})$ is a complex function, $\sigma_{\pm}= \sigma_x \pm i\sigma_y$, and $\sigma_{i}$ ($i=x$, $y$) are the Pauli matrices. Here, we ignore the the kinetic term in Eq. (\ref{eqH}). In the absence of SOC, there are only majority-spin states across the Fermi level. The decoupling between the spin and orbital species indicates that the vdW germanene/Cr$_2$Ge$_2$Te$_6$ heterostructure can be considered as a spinless ferromagnet. In this case, all crystalline symmetries as well as the $\mathcal{T}$-symmetry are preserved for each spin channel. At the $\mathcal{T}$-invariant $\Gamma$ point, the $k\cdot p$ Hamiltonian is subjected to the product of $C_{3}$ little group and $\mathcal{T}$, which constrains Eq. (\ref{eqH}) as
\begin{equation} \label{HconT}
{C}_{3}\mathcal{T}\mathcal{H}(\mathbf{k})\mathcal{T}^{-1}{C}_{3}^{-1}=\mathcal{H}[(\mathbf{R}_{3}\mathcal{T})\mathbf{k}],
\end{equation}
where $\mathbf{R}_{3}$ is a $2\times 2$ rotational matrix of $C_3$. In the spinless-like system, we have $\mathcal{T}^2=1$ and thus the $\mathcal{T}$-operator can be represented as $\mathcal{T}=K$ with a complex conjugate operator $K$. As a result, the constraint of Eq. (\ref{HconT}) gives (see the SM \cite{SM})
\begin{equation}\label{dft}
e^{i2\pi/3}d(k_{+}, k_{-})=f(k_{+}e^{-i\pi/3}, k_{-}e^{i\pi/3}),
\end{equation}
where $k_{\pm}=k_x\pm ik_{y}$. Then, we can expand Eq. (\ref{dft}) and obtain the symmetry-allowed expressions to the lowest orders as
\begin{equation}\label{AG}
f(\mathbf{k})=a_{+}k_+^{2}+a_{-}k_-^{2},
\end{equation}
which uncovers that low-energy excitations around the $\Gamma$ point are indeed quadratic in the $k_x$-$k_y$ plane.

\begin{figure}
	\centering
	\includegraphics[scale=0.47]{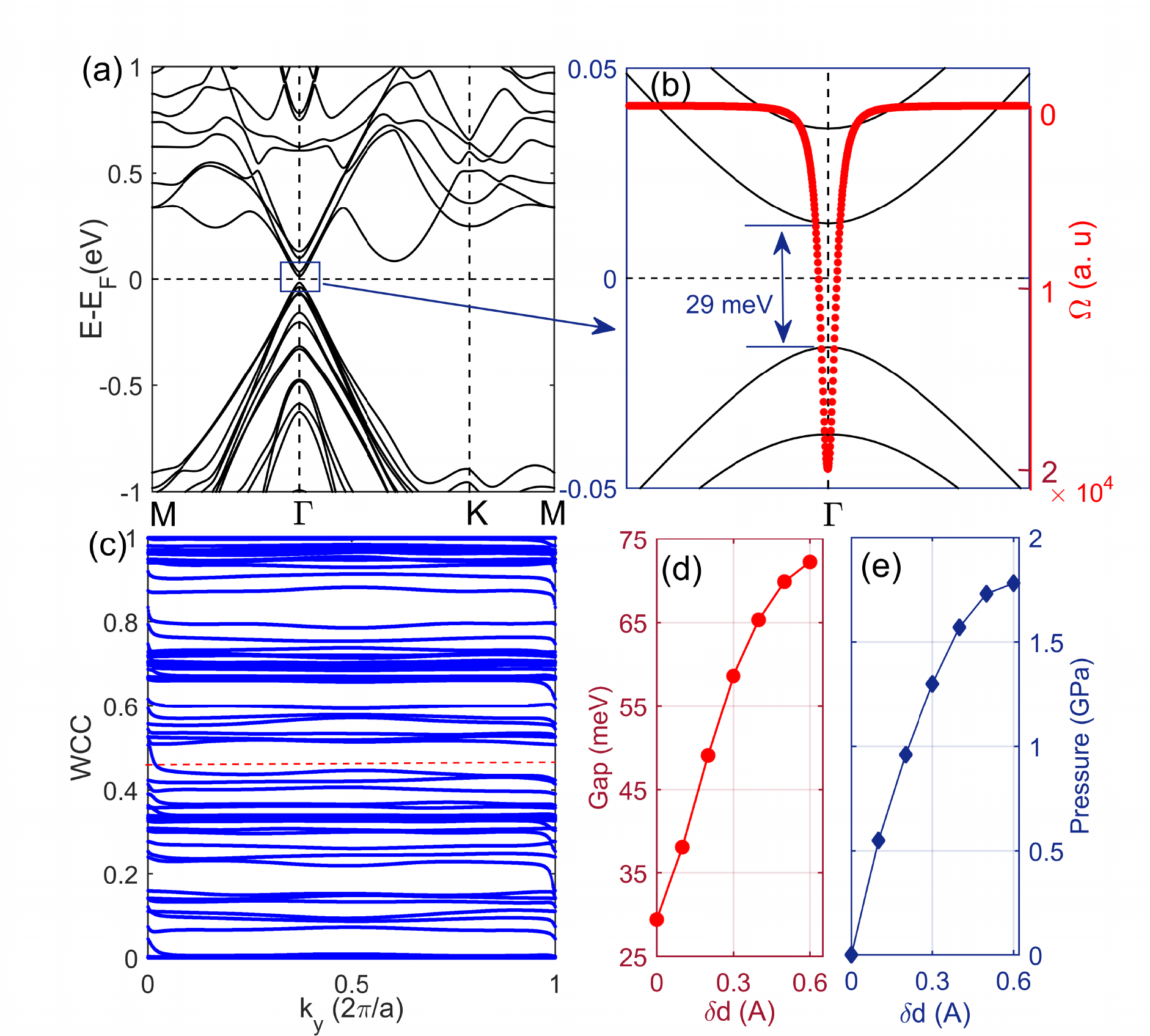}
	\caption{(a) The electronic band structure of the vdW germanene/Cr$_2$Ge$_2$Te$_6$ heterostructure with magnetization perpendicular to the 2D plane. (b) The enlarged view of (a) near the $\Gamma$ point, showing a nontrivial band gap of $\sim$29 meV. The calculated Berry curvature illustrated as the red-circles. (c) Evolution of WCCs as a function of $k_y$. (d) Nontrivial band gap as a function of reduction of interlayer distance. (e) External pressure as a function of reduction of interlayer distance.
\label{figure3}}
\end{figure}

In the presence of SOC, the coupling between the spin and orbital species causes spontaneous breaking of spin rotational symmetry, and thus the $\mathcal{T}$-symmetry is also removed. In this case, the magnetic space group strongly influences the topological features. To determine the magnetization direction, we have carried out total energy calculations with different magnetic configurations, including magnetization along out-of-plane and in-plane directions. Our results confirm that the easy axis is perpendicular to the 2D plane, which agrees well with the magnetic groundstate properties of Cr$_2$Ge$_2$Te$_6$ in experiments \cite{Naturecrgete}. The electronic band structures with magnetization perpendicular to the 2D plane are shown in Figs. \ref{figure3}(a) and \ref{figure3}(b). The figures show that the vdW germanene/Cr$_2$Ge$_2$Te$_6$ heterostructure converts into a insulator with a sizable band gap of 29 meV. It is worth noting that this band gap is larger than that of germanene. Since the Cr $3d$ and Te $5p$ orbitals host the greater SOC strength than the Ge $4p$ states, and thus the interlayer coupling and hybridization can enlarge the band gap. To confirm whether this band gap opened by SOC is nontrivial, we calculate the Berry curvature as implemented in the Wannier90 code \cite{Mostofi2008}. In Fig. \ref{figure3}(b), the calculated Berry curvature, illustrated as the red-dots, shows a peak near the $\Gamma$ point. We integrate the Berry curvature over the states below the Fermi level, and obtain the nonzero Chern number $\mathcal{C=}-1$. This implies that the QAH state is present in the vdW germanene/Cr$_2$Ge$_2$Te$_6$ heterostructure. We also calculate the evolution of Wannier charge centers (WCCs) to further verify the nontrivial properties. The WCCs are calculated using Z2PACK package\cite{PhysRevB.95.075146}, in which the Wilson loop method is employed  \cite{Yu2011}. As shown in Fig. \ref{figure3}(c), we can see that the referenced line can always cross the WCC line once, confirming that this system is indeed a nontrivial QAH insulator. In addition, the energy differences between out-of-plane and in-plane magnetic configurations are smaller than $\sim$1 meV. This implies that there may be the possibility of the topological phase transition between the QAH and 2D Weyl semimetallic states by applying an external field \cite{PhysRevB.100.064408}.

Due to the weak interaction of vdW heterostructures, the interlayer distance can be easily reduced by a vertical external pressure. The applied external pressure at the interlayer distance $d$ is calculated by $P=\delta E/(S \delta d)$, where $\delta E=E(d)-E(d_{0})$ represents the energy difference between the compressed $E(d)$ and equilibrium $E(d_{0})$ structures, $\delta d = |d-d_0|$, and $S$ is the area per unit cell of the heterostructure. The reduction of the interlayer distance can enhance the hybridization between the germanene layer and the Cr$_2$Ge$_2$Te$_6$ layer, which may further enlarge the nontrivial band gap. The corresponding electronic band structures without and with SOC are given in the SM \cite{SM}. As expected, we can see that the nontrivial band gaps are monofonically increased with reducing the interlayer distance in Fig. \ref{figure3}(e).  When the vertical external pressure is 1.8 GPa (i.e., $\delta d= 0.6$ {\AA}), the nontrivial band gap can be considerably up to $\sim$72 meV [see Figs. \ref{figure3}(d) and \ref{figure3}(e)]. Besides, we also find that the germanene/Cr$_2$Ge$_2$Te$_6$ heterostructure will transform to a trivial phase when $\delta d>0.7$ {\AA}.

\begin{figure}
	\centering
	\includegraphics[scale=0.49]{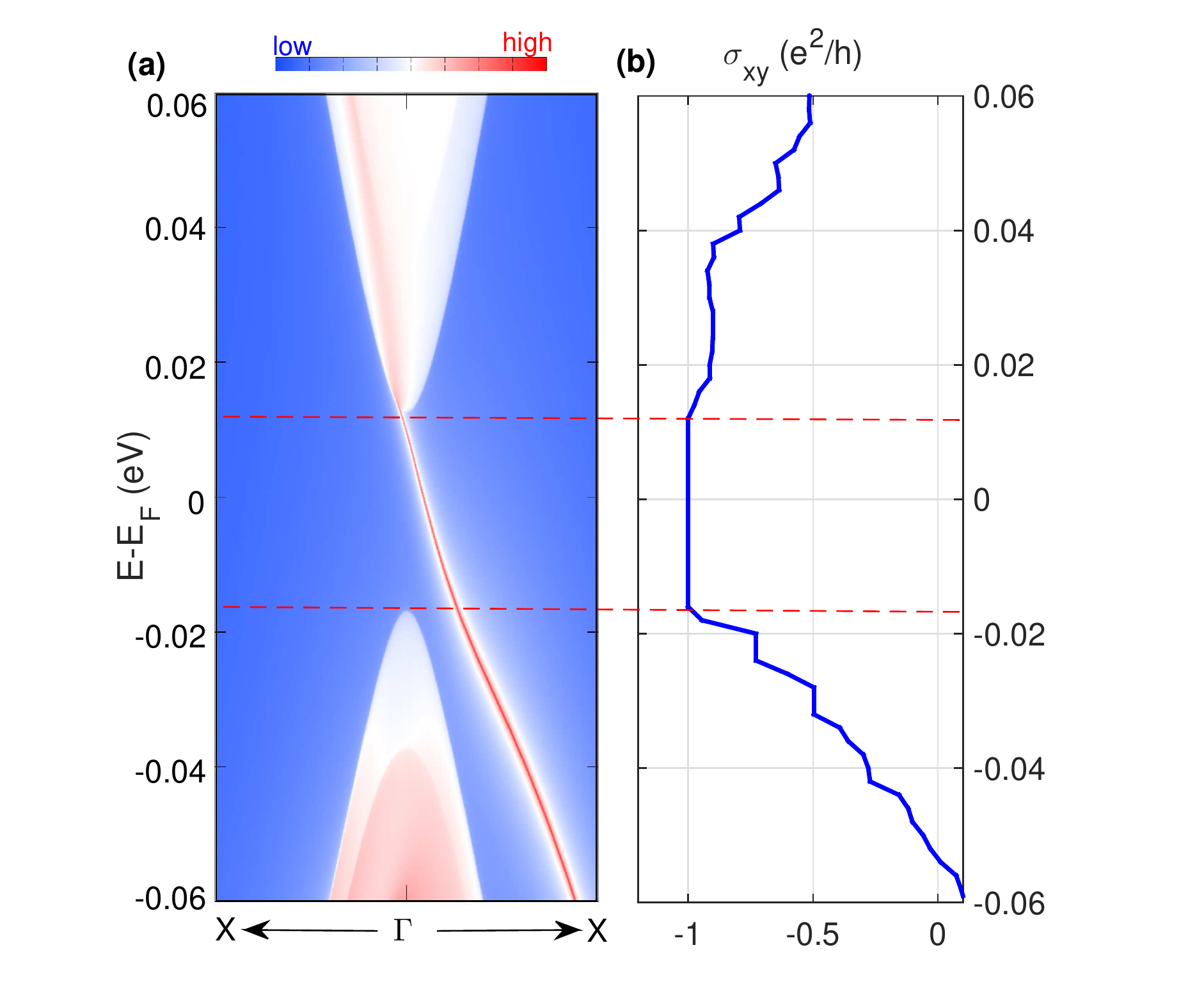}
	\caption{(a) The calculated LDOS of a semi-infinite ribbon of the germanene/Cr$_2$Ge$_2$Te$_6$ heterostructure. (b) The intrinsic Hall conductance inside the nontrivial band gap exactly quantized to $-e^2/h$.
\label{figure4}}
\end{figure}

The QAH insulator with the nonzero Chern number $\mathcal{C}=-1$ possesses the chiral edge states and quantized Hall conductance. To illustrate these topologically protected features, we construct a tight-binding (TB) Hamiltonian based on maximally localized Wannier functions methods \cite{Mostofi2008}. The local density of states (LDOS) are calculated by the iterative Green's method \cite{Sancho1984,WU2017}, and the anomalous Hall conductance are calculated from the Kubo formula with the TB Hamiltonian. The LDOS of a semi-infinite ribbon of the germanene/Cr$_2$Ge$_2$Te$_6$ heterostructure and Hall conductance are respectively shown in Fig. \ref{figure4}(a) and \ref{figure4}(b). As expected, the chiral edge states connecting the valence and conduction bands are visible. The intrinsic Hall conductance $\sigma_{xy}$ is exactly quantized to $-e^2/h$ inside the nontrivial band gap.

In conclusion, we demonstrate by using first-principles calculations and effective model analysis that the robustly intrinsic QAH effect can be realized in the vdW heterostructure germanene/Cr$_2$Ge$_2$Te$_6$. We show that hybridization and charge transfer occur between the germanene and Cr$_2$Ge$_2$Te$_6$ layers. The proximity effect gives rise to a quadratic nodal point in band structures in the absence of SOC. When SOC is considered and the magnetization is aligned along the easy axis, the quadratic nodal point is opened, driving this heterostructure to be an QAH insulator with a sizable band gap of 29 meV. In addition, we also find that the nontrivial band gap can be tunable up to 72 meV by applying an external pressure. The chiral edge states and quantized Hall conductance have been investigated, which can be easily detectable. Considering that Cr$_2$Ge$_2$Te$_6$ is a nearly ideal 2D ferromagnet, the vdW heterostructure germanene/Cr$_2$Ge$_2$Te$_6$ with high thermodynamic stability is expected to its synthesis in the laboratory. Therefore, our work provides a reliable platform to observe intrinsic QAH effects and design topological devices in futures.

This work was supported by the National Natural Science Foundation of China (NSFC, Grants No. 11974062, No. 11704177, and 11947406), the Chongqing National Natural Science Foundation (Grants No. cstc2019jcyj-msxmX0563), and the Fundamental Research Funds for the Central Universities of China (Grant No. 2019CDXYWL0029)\\
~~~\\

%

\end{document}